\pdfoutput=1

\documentclass[Afour,sagev,times]{sagej}

\usepackage{moreverb,url}

\usepackage[colorlinks,bookmarksopen,bookmarksnumbered,citecolor=red,urlcolor=red]{hyperref}

\newcommand\BibTeX{{\rmfamily B\kern-.05em \textsc{i\kern-.025em b}\kern-.08em
T\kern-.1667em\lower.7ex\hbox{E}\kern-.125emX}}

\def\volumeyear{2016}

\begin{document}

\runninghead{Kuroda et al.}

\title{Apparatus for High-Precision %
Angle-Resolved Reflection Spectroscopy %
in the Mid-Infrared Region}

\author{%
Takashi~Kuroda\affilnum{1,2}, %
Siti~Chalimah\affilnum{1,2}, %
Yuanzhao~Yao\affilnum{1}, %
Naoki~Ikeda\affilnum{1}, %
Yoshimasa~Sugimoto\affilnum{1}, %
and Kazuaki~Sakoda\affilnum{1} %
}

\affiliation{%
\affilnum{1}National Institute for Materials Science, Tsukuba, Japan\\
\affilnum{2}Graduate School of Engineering, Kyushu University, Japan%
}

\corrauth{Takashi~Kuroda, %
National Institute for Materials Science, 1-1 Namiki, Tsukuba 305-0044, Japan%
}

\email{kuroda.takashi@nims.go.jp}

\begin{abstract}
Fourier transform (FT) spectroscopy is a versatile technique for studying the infrared (IR) optical response of \, solid-, liquid-, and gas-phase samples. In standard FT-IR spectrometers, a light beam passing through a Michelson interferometer is focused onto a sample with condenser optics. This design enables us to examine relatively small samples% with a sufficient throughput
, but the large solid angle of the focused infrared beam makes it difficult to analyze angle-dependent characteristics. Here we design and construct a high-precision angle-resolved reflection setup compatible with a commercial FT-IR spectrometer. Our setup converts the focused beam into an achromatically collimated beam with an angle dispersion as high as %$\mathsf{\pm0.4^{\circ}}$.
$\mathsf{0.25^{\circ}}$. The setup also permits us to scan the incident angle over $\mathsf{\sim8^{\circ}}$ across zero (normal incidence). The beam diameter can be reduced to $\mathsf{\sim1}$~mm, which is limited by the sensitivity of an HgCdTe detector. The small-footprint apparatus is easily installed in an FT-IR sample chamber. As a demonstration of the capability of our reflection setup we measure the angle-dependent mid-infrared reflectance of two-dimensional photonic crystal slabs and determine the in-plane dispersion relation in the vicinity of the $\mathsf{\Gamma}$ point in momentum space. We observe the formation of photonic Dirac cones, i.e., linear dispersions with an accidental degeneracy at $\mathsf{\Gamma}$, in an ideally designed sample. Our apparatus is %expected to serve as a useful tool
useful for characterizing various systems that have a strong in-plane anisotropy, including photonic crystal waveguides, plasmonic metasurfaces, and molecular crystalline films. 
\end{abstract}


\maketitle

\noindent
\textcolor{blue}{%
This is the version of the article accepted for publication in \textit{Applied Spectroscopy}. The final published version will be available open access from the journal's site. %
}%
\section{Introduction}
The Fourier transform infrared (FT-IR) spectrometer is a ubiquitous spectroscopic tool used in a wide range of material research \cite{Bell,Griffiths_and_Haseth}. Standard FT-IR spectrometers have two optical configurations; transmission and reflection. Reflection measurement allows us to study optically thick and highly absorbing samples whose transmission spectra are difficult to obtain. Nevertheless, a rigorous Kramers-Kronig analysis of specular reflection spectra provides both the real and imaginary refractive indices of materials \cite{Harbecke,Ohta_and_Ishida,Kocak}, the same information gained by transmission measurements of carefully prepared optically thin samples. Moreover, attenuated total reflection (ATR) techniques are used to measure the spectra of small amounts of samples \cite{Lewis00,Patterson_06,Dallen_07}. The combination of an FT-IR with a Cassegrain-type objective lens or tip-enhanced near-field optic is capable of infrared imaging with a sub-micrometer resolution \cite{Knoll_and_Keilmann,Taubner,Furstenberg_RSI06}. Such spatial resolution techniques, however, could fail to analyze the angular-dependent optical response inherent in anisotropic samples. 

The loss of angular resolution is also a problem with standard FT-IR architecture, where a thick beam from a Michelson interferometer is tightly focused on a sample. The angle of view at the sample position can reach $2\theta \sim 20 ^{\circ}$ (numerical aperture (NA) of $\sim 0.17$), which thus easily masks (or even eliminates) the angle-dependent fine resonance in infrared spectra. The optical design differs from that of ultraviolet-visible (UV-Vis) spectrometers, where a roughly collimated (unfocused) beam is transmitted through the sample. Thanks to the availability of large-area photodetectors in the UV-Vis wavelength regions, a collimated beam passing through the sample chamber is directly coupled to the detector. In contrast, infrared detectors have much smaller apertures and thus require a focusing design for FT-IR. 

Two kinds of commercial products are currently available that make it possible to measure angle-resolved infrared spectra. Variable angle reflection attachments are useful FT-IR accessories%\cite{example1}. 
\endnote{Examples for variable angle reflection attachments to FT-IR: Seagull (Harrick Scientific Products) or VeeMAX III (Pike Technologies).}%
. They can be operated over broad incident angles while preserving optical alignment. However, the incident beam is designed to focus on the sample position, and the dispersion of the incident angles often exceeds 10$^\circ$. Hence, it is difficult to study samples that have a strong angular dependence, as the spectral signatures are masked due to the convolution of different angle-dependent spectra. To achieve sub-$1 ^\circ$ angle resolution we develop an alternative setup that converts the focused infrared beam (inside an FT-IR sample room) into a collimated beam using parabolic collimators. %
Spectroscopic ellipsometers are also devices that take advantage of angle-resolved reflection. Some advanced systems are combined with FT-IR%\cite{example2,Ferrieu_RSI89,Canillas_RSI93}
\endnote{Examples for infrared spectroscopic ellipsometers: IR-VASE Mark II (J.~A.~Woolam) or Sendira (Sentech).}%
\cite{Ferrieu_RSI89,Canillas_RSI93}. However, ellipsometers generally use a strongly oblique incidence, and prevent us from measuring normal incidence spectra. %

ATR-FT-IR spectroscopic imaging techniques are capable of increasing spatial resolutions beyond free-space diffraction limited values thanks to the use of high index materials for internal reflection elements. In addition, ATR imaging with variable angles of incidence makes it possible to obtain depth profiles, where the angles of incidence are restricted and controlled by inserting carefully designed apertures in commercially available macro ATR accessories \cite{Kazarian_ApplSpctr07,Kazarian_ApplSpctr08,Kazarian_Langmuir10} 
and Cassegrain microobjectives \cite{Kazarian_ApplSpctr15,Kazarian_Analyst19}. However, these imaging techniques essentially require the finite distributions of wavevectors (angles of incidence), and the severe light blocking (the use of small apertures) would be needed to achieve sub-$1^{\circ}$ angle resolution. To obtain a good angle resolution with sufficient throughput, we adopt a design that uses a collimated beam rather than a focused beam, although the imaging capability is lost. 

\begin{figure}
\centering
\includegraphics[height=5cm]{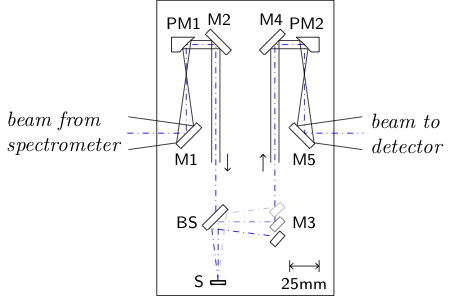}
\caption{\label{fig_scheme} Optical diagram of the angle-resolved reflection setup. M1, M2, M4, and M5 plane mirrors, PM1 parabolic collimator, BS beam splitter, S rotatable sample holder, M3 movable plane mirror, and PM2 parabolic condenser (identical to PM1). %
The (blue) dash dotted line indicates the optical axis. All the optics are arranged on a 150~mm $\times$ 250~mm breadboard depicted by the rectangular frame. }
\end{figure}

In this work we develop an angle-resolved mid-infrared reflection setup with two design objectives. The first objective is to achieve a good collimation for the incident beam whose incident angle is tunable across zero, i.e., the normal incidence angle. The setup thus utilizes a parabolic collimator/condenser pair, and an infrared compatible beamsplitter, which transmits the beam to the sample, and reflects it to the detector. The beam shift associated with sample rotation is fully compensated for in the setup. 
%All the optic and mechanical devices used for the setup are chosen from commercial products. 
The second design objective is a small footprint, which means that the setup can smoothly be installed in the sample room of a standard FT-IR system. The modular design facilitates the flexibility of entire FT-IR systems in laboratories. As a demonstration of the capability of the reflection setup, we measure the angle-dependent mid-infrared reflection spectra of two-dimensional photonic crystal (PC) slabs, which are fabricated on silicon-on-insulator (SOI) substrates \cite{Yao_OEX20}. 

\section{Design and Installation}
We use FT/IR6800 (Jasco), which has a maximum resolution of 0.07~cm$^{-1}$, as a base FT-IR spectrometer. It incorporates a liquid nitrogen cooled mercury cadmium telluride (MCT) detector, which has good sensitivity from 650 to 12000~cm$^{-1}$. 

Figure~\ref{fig_scheme} is an optical diagram of the reflection setup. Infrared light from the spectrometer is reflected by a flat steering mirror M1 to an off-axial parabolic mirror PM1 that has a focal length of 25.4~mm (Thorlabs MPD019-M01). PM1 collimates the beam and sends it to another steering mirror M2. The beam is then directed to a 50:50 calcium fluoride beamsplitter (BS; Thorlabs BSW510 for $\lambda=$~2\,-\,8~$\mu$m). The beam transmitted through the BS is incident on a sample, which is mounted on a rotary stage. The sample reflects the beam and returns it to the BS, which reflects it to a movable mirror M3. The motion of M3 fully compensates for the beam shift with changing the incident angle: When the sample is rotated by $\theta$, M3 is rotated by $-\theta$, and translated along the optical axis so that the beam follows the same path. Then, the beam passes through M4, PM2 (identical to PM1) and M5 to the instrument detector. 

\begin{figure}[t]
\centering
\includegraphics[height=5cm]{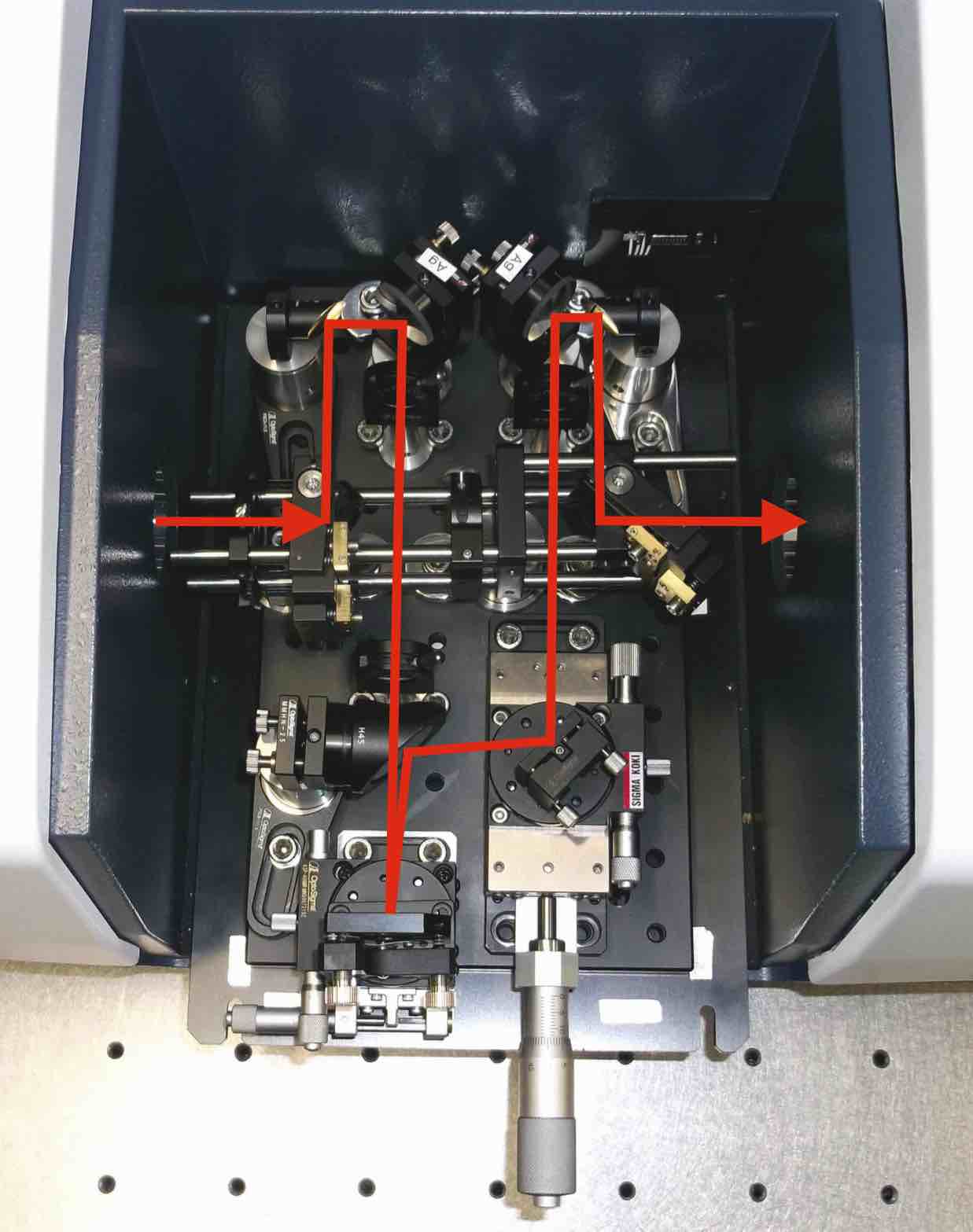}
\caption{\label{fig_photo} Photograph of the angle-resolved reflection setup, which is installed in FT/IR6800 (Jasco). The red line highlights the infrared beam path. }
\end{figure}

An iris pair is inserted in between M2 and BS to reduce the beam diameter down to 1~mm, which is limited by the MCT sensitivity. All the optics are assembled on a 150~mm$\times$250~mm aluminum breadboard, which can be placed directly in the FT-IR sample chamber (see the setup image in Fig.~\ref{fig_photo}). %

Proper optics alignment is essential to obtain a good angular resolution. The use of a visible light guide makes the alignment rather easy. In particular, we use a fiber coupled semiconductor laser as a point light source, and a fiber coupled photodiode as a small aperture detector. Before placing the optics, we carefully construct an alignment configuration with a visible light path (from source to detector), which imitates the infrared light path in FT-IR. Note that we prepare this alignment configuration outside the FT-IR system. We install the breadboard there and adjust all the optics positions using visible light. After achieving a good condition, we move the breadboard (with all the optics in the correct place) to FT-IR, %
and carry out experiments. Note that the reflection setup reported in our earlier work utilized plano-concave lenses as a collimator/condenser pair \cite{Yao_OEX20}. Here we replace the concave lenses with parabolic mirrors (PM1 and PM2), and remove chromatic aberrations. Hence, optical alignment using a visible laser, which has a wavelength far from that of IR light, is an efficient way to achieve optimal conditions. 

\begin{figure}
\centering
\includegraphics[width=5.5cm]{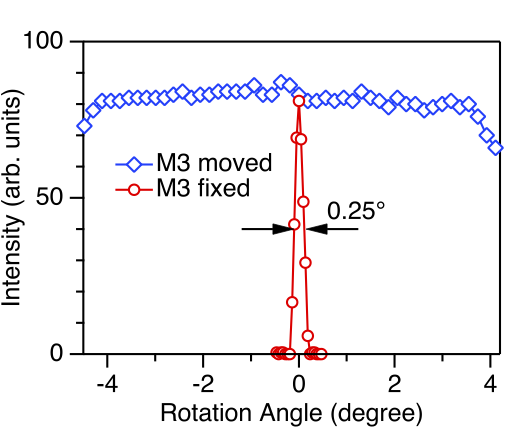}
\caption{\label{angleRsltn} Mirror reflection intensity as a function of rotation angle. For the red circles we rotate the mirror holder, but we do not move the compensator (M3). For the blue diamonds we move the compensator while simultaneously rotating the mirror. %The data are normalized to their maximum values. 
}
\end{figure}
\section{Performance test}
\subsection{Estimation of angle resolution}
%\section{Result and Discussion}
%\subsection{Performance test}
Figure~\ref{angleRsltn} shows the reflection intensity of a silver mirror mounted on the sample holder as a function of rotation angle. Here we use the mirror as a reflection standard and define the intensity as the spectrally integrated intensity of the FT-IR output. The red circles show the intensity variation when the compensation optic (M3) does not move. The intensity peaks at zero, and then suddenly decreases when the mirror is rotated. Hence, the reflection setup is highly sensitive to the rotation (i.e., incident) angle. The full width at half maximum of the measured peak is %$\sim 0.8^\circ$,
evaluated to be $0.25^\circ$, which thus indicates the angle resolution of the setup. The blue diamonds show the intensity variation, while we adequately move M3 to compensate for the beam shift associated with mirror rotation. The measured intensity remains almost constant even when the incident angle is changed, which ensures the stability of the setup. Currently, the tunable range of the incident angle is $\sim 8^\circ$, which is limited by the BS size (25.4~mm in diameter) and the M3 translator travel (25~mm). The tunable range could be extended by making minor changes. 

\begin{figure}
\centering
\includegraphics[height=5.5cm]{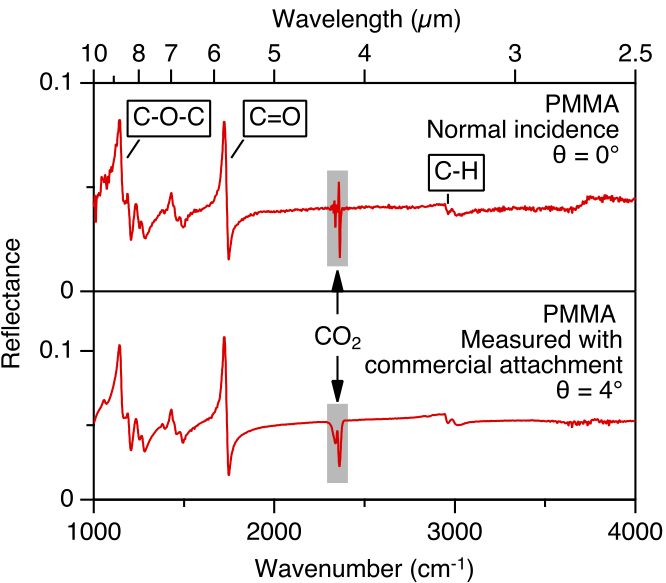}
\caption{\label{PMMA_spctr} Reflection spectra of a bulk PMMA measured using our reflection setup at normal incidence (top), and a commercially available variable-angle reflection attachment, where the incident angle is set at $4^{\circ}$ (bottom). Spectral noises associated with the absorption of CO$_2$ are indicated by the arrows. The assignments of several vibrational modes in PMMA are also indicated. %
}
\end{figure}

The top panel in Fig.~\ref{PMMA_spctr} shows the reflection spectrum of a bulk polymethyl methacrylate (PMMA) at normal incidence ($\theta=0^{\circ}$). The spectrum indicates a variety of resonance peaks, which resemble the first derivative curves of symmetric absorption peaks assigned to different vibrational modes. The bottom panel shows the spectrum of the same PMMA measured using a commercially available variable-angle reflection accessory (Harrick Scientific, Seagull) at near normal incidence ($\theta=4^{\circ}$). The two spectra are almost identical thanks to the optical isotropy of PMMA. The homogeneous sample exhibits a spectrum that is not very sensitive to the incident angles. Nevertheless, the measured spectral reproducibility ensures the accuracy of our hand-made setup. %

\subsection{Mid-IR characterization of photonic crystal slabs}
We apply the angle-resolved reflection setup to the characterization of PC slab waveguides based on an SOI, which contains a top silicon layer with a thickness of 400~nm. The target PC design is a square lattice of air holes with a lattice constant of 2.27~$\mu$m, a hole radius (R) of 526~nm, and a depth (d) of 216~nm. These PC parameters lead to the formation of photonic Dirac cones, i.e., accidental degeneracy in the waveguide modes at the $\Gamma$ point in momentum space \cite{Sakoda_OEX12_1,Sakoda_OEX12_2}. The samples are fabricated using a combination of electron beam lithography and reactive ion beam etching techniques. The design and fabrication are detailed elsewhere \cite{Yao_OEX20}. 

\begin{figure}
\centering
\includegraphics[height=5.5cm]{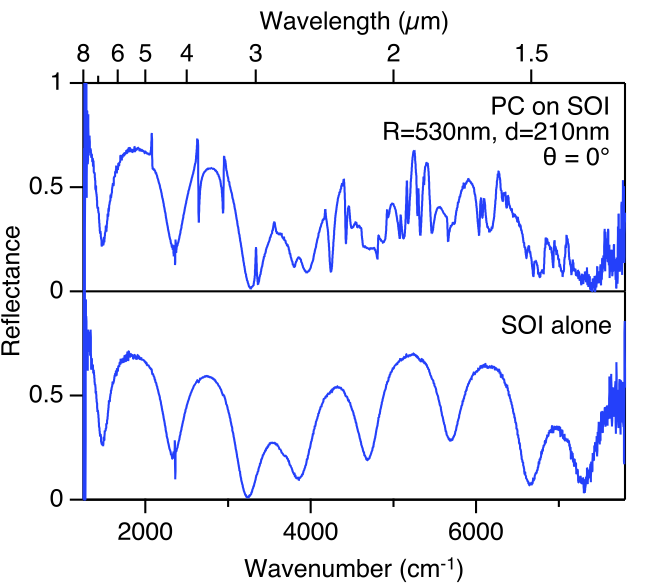}
\caption{\label{normalIncdncwSOI} Mid-IR reflectance of an SOI PC slab (R = 530 nm and d = 210 nm) at normal incidence. The spectrum of an unprocessed SOI is also shown at the bottom. }
\end{figure}
Figure~\ref{normalIncdncwSOI} shows the normal incidence reflection spectrum of a PC sample together with that of an unprocessed SOI. Both spectra exhibit a slowly varying profile similar to a modulated sinusoid. Additionally, the PC sample spectrum has various small peaks. Thus, the slowly varying components appear due to Fabry-Perot interference occurring in the SOI, and the additional peaks arise due to resonant coupling between the incident light and the waveguide modes. Since the wavevector of light at normal incidence does not have a projection onto the slab layer, the incident light is only coupled to the waveguide mode at the $\Gamma$ point ($k_{||} = 0$). Consequently, the ensemble of sharp spectral lines represents the mode distribution at $\Gamma$ as a function of frequency. 

\begin{figure}
\centering
\includegraphics[height=6cm]{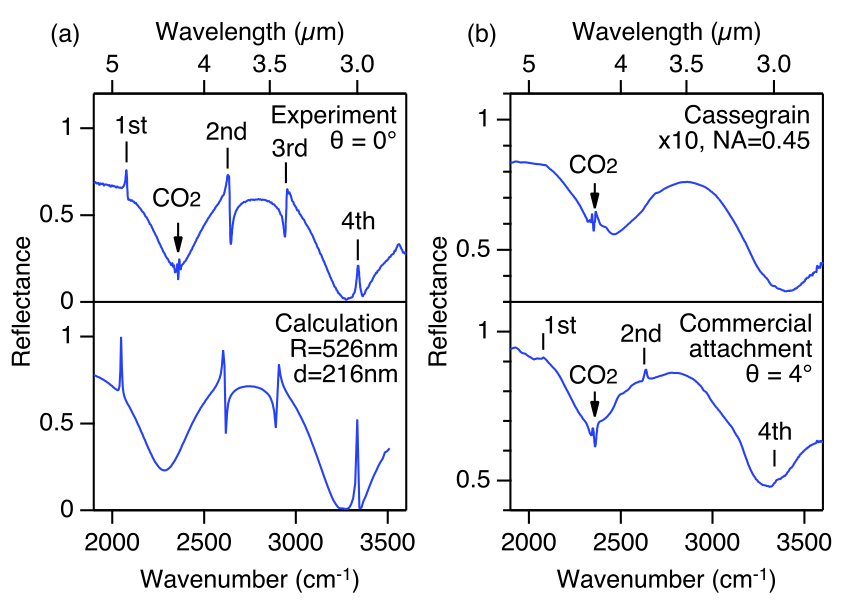}
\caption{\label{normalIncdnc_cmprsn} (a) Comparison of measured (top) and calculated (bottom) reflection spectra at normal incidence. We assume R~=~526~nm and d~=~216~nm for our calculation. Spectral noises associated with the absorption of CO$_2$ are indicated by the arrows. (b) Reflection spectra measured using a Cassegrain objective (top), and a commercial variable-angle reflection attachment, where the incident angle is set to 4$^\circ$ (bottom). }
\end{figure}
The top panel in Fig.~\ref{normalIncdnc_cmprsn}(a) shows an expanded view of the normal incident spectrum at wavenumbers around 2750~cm$^{-1}$, i.e., the low frequency side of the spectrum in Fig.~\ref{normalIncdncwSOI}. Four resonant peaks are found in this spectral range (highlighted by the vertical lines), and Dirac cones are expected to appear at the second lowest energy peak, as discussed later. The bottom panel in Fig.~\ref{normalIncdnc_cmprsn}(a) shows the finite element calculation result, which perfectly reproduces the measured spectrum. Figure~\ref{normalIncdnc_cmprsn}(b) shows the spectra measured using common reflection accessories of FT-IR: In the top panel we use an infrared microscope with a Cassegrain optic with NA~$=0.45$ ($\times$10 magnification, finite conjugate design), and in the bottom panel we use a standard variable-angle reflection attachment (Harrick Scientific, Seagull). Neither spectrum exhibits clear resonance peaks, as they are masked due to the convolution of different spectra with distributed incident angles. The correct spectra are thus available for a sufficiently high angle resolution, which we have successfully achieved in this work. 

\begin{figure}
\centering
\includegraphics[height=6cm]{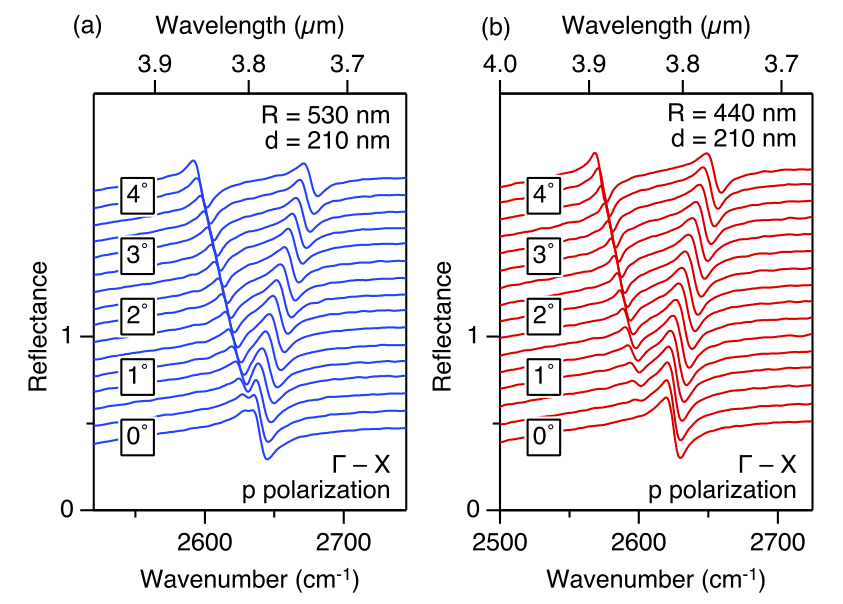}
\caption{\label{angleDpndnc} Incident angle dependent reflection spectra for PC slabs with (a) R = 530 nm, and (b) R = 440 nm. The spectra are arranged from bottom to top, $\theta =$~0 to 4.38$^\circ$ in 0.292$^\circ$ steps. The incident beam is p-polarized and tilted towards the [100] in-plane axis of the square lattice sample. }
\end{figure}
Figures~\ref{angleDpndnc}(a) and \ref{angleDpndnc}(b) show the reflection spectra with different incident angles $\theta$ for samples with different hole radii. %with a step of 0.292$^\circ$. 
Here, the incident beam is tilted in the incident plane towards the [100] axis of the square PC lattice. Accordingly, the wavevector is moved between $\Gamma$ and X in the first Brillouin zone. The sample with R~=~530~nm (Fig.~\ref{angleDpndnc}(a)) shows that the single peak at $\theta=0$ splits into two peaks almost linearly with $|\theta|$. This is the signature of Dirac cones formed at $\Gamma$. In contrast, the sample with R~=~440~nm (Fig.~\ref{angleDpndnc}(b)) exhibits a roughly parabolic energy shift in the vicinity of $\theta=0$. In parallel, another new peak appears at finite $\theta$ values. The absence of the new peak at $\theta=0$ reflects the optical selection rule of the PC mode \cite{Yao_OEX20}. Hence, no degeneracy is observed at $\theta=0$ in this sample. Thus, we confirm the formation of Dirac cones in a sample that has a PC parameter close to the theoretically predicted value (R~=~526~nm). 

\section{Conclusions}
We developed an angle-resolved reflection setup compatible with a standard FT-IR spectrometer. The angle resolution reached %$\pm0.4^\circ$, 
$0.25^\circ$, which is not available with commercial products. The setup makes it possible to measure normal incidence spectra, and to clarify the in-plane band dispersion of photonic crystal waveguides in the vicinity of the $\Gamma$ point. Thus, one potential application of the setup is the study of PC surface-emitting lasers, which employ laser action at the $\Gamma$ point. Photonic band characterization has already been realized by using angle-dependent sub-threshold luminescence measurement \cite{Williams_PTL12,Taylor_JSTQE12}. However, the technique cannot be applied to mid-infrared lasers due to the limited sensitivity of infrared detectors. Our setup is expected to serve as a tool with which to clarify the in-plane waveguide dispersion, and confirm vertical emission even in mid-infrared regions. 

\begin{acks}
The authors thank Ms. Miwako Kobayashi and Mr. Masao Ohya (Jasco Corp.) for their technical assistance, Hiromi Koyama and Takaaki Mano for helpful discussions. SC gratefully acknowledges support from the NIMS Graduate Assistantship program.  
\end{acks}

\begin{dci}
The authors declare that there is no conflict of interest.
\end{dci}

\begin{funding}
This work was supported by Innovative Science and Technology Initiative for Security, Grant Number JPJ004596, ATLA, Japan. 
\end{funding}

\theendnotes

\bibliographystyle{SageV}
\bibliography{anglRslvd.bib}

\begin{thebibliography}{10}
\providecommand{\url}[1]{\texttt{#1}}
\providecommand{\urlprefix}{URL }
\expandafter\ifx\csname urlstyle\endcsname\relax
  \providecommand{\doi}[1]{DOI:\discretionary{}{}{}#1}\else
  \providecommand{\doi}{DOI:\discretionary{}{}{}\begingroup
  \urlstyle{rm}\Url}\fi
\providecommand{\eprint}[2][]{\url{#2}}

\bibitem{Bell}
Bell RJ.
\newblock \emph{Introductory Fourier Transform Spectroscopy}.
\newblock Academic Press, 1972.
\newblock ISBN 978-0-12-085150-8.
\newblock \doi{10.1016/B978-0-12-085150-8.X5001-3}.
\newblock \urlprefix\url{https://doi.org/10.1016/B978-0-12-085150-8.X5001-3}.

\bibitem{Griffiths_and_Haseth}
Griffiths PR and de~Haseth JA.
\newblock \emph{Fourier Transform Infrared Spectrometry}.
\newblock 2nd. ed. John Wiley {\&} Sons, 2006.
\newblock ISBN 9780470106310.
\newblock \doi{10.1002/047010631X}.
\newblock \urlprefix\url{https://doi.org/10.1002/047010631X}.

\bibitem{Harbecke}
Harbecke B.
\newblock {``Application of {F}ourier's Allied Integrals to the
  {Kramers-Kronig} Transformation of Reflectance Data''}.
\newblock \emph{Appl~Phys~A} 1986; 40(3): 151--158.
\newblock \doi{10.1007/BF00617396}.
\newblock \urlprefix\url{https://doi.org/10.1007/BF00617396}.

\bibitem{Ohta_and_Ishida}
Ohta K and Ishida H.
\newblock {``Comparison among Several Numerical Integration Methods for
  {Kramers-Kronig} Transformation''}.
\newblock \emph{Appl~Spectrosc} 1988; 42(6): 952--957.
\newblock \doi{10.1366/0003702884430380}.
\newblock \urlprefix\url{https://doi.org/10.1366/0003702884430380}.

\bibitem{Kocak}
Kocak A, Berets SL, Milosevic V et~al.
\newblock {``Using the {Kramers-Kronig} Method to Determine Optical Constants
  and Evaluating its Suitability as a Linear Transform for Near-Normal
  Front-Surface Reflectance Spectra''}.
\newblock \emph{Appl~Spectrosc} 2006; 60(9): 1004--1007.
\newblock \doi{10.1366/000370206778397443}.
\newblock \urlprefix\url{https://doi.org/10.1366/000370206778397443}.

\bibitem{Lewis00}
Lewis LL and Sommer AJ.
\newblock {``Attenuated Total Internal Reflection Infrared Mapping
  Microspectroscopy of Soft Materials''}.
\newblock \emph{Appl~Spectrosc} 2000; 54(2): 324--330.
\newblock \doi{10.1366/0003702001949294}.
\newblock \urlprefix\url{https://doi.org/10.1366/0003702001949294}.

\bibitem{Patterson_06}
Patterson BM and Havrilla GJ.
\newblock {``Attenuated Total Internal Reflection Infrared Microspectroscopic
  Imaging Using a Large-Radius Germanium Internal Reflection Element and a
  Linear Array Detector''}.
\newblock \emph{Appl~Spectrosc} 2006; 60(11): 1256--1266.
\newblock \doi{10.1366/000370206778998941}.
\newblock \urlprefix\url{https://doi.org/10.1366/000370206778998941}.

\bibitem{Dallen_07}
van Dalen G, Heussen PCM, Adel RD et~al.
\newblock {``Attenuated Total Internal Reflection Infrared Microscopy of
  Multilayer Plastic Packaging Foils''}.
\newblock \emph{Appl~Spectrosc} 2007; 61(6): 593--602.
\newblock \doi{10.1366/000370207781269738}.
\newblock \urlprefix\url{https://doi.org/10.1366/000370207781269738}.

\bibitem{Knoll_and_Keilmann}
Knoll B and Keilmann F.
\newblock {``Near-Field Probing of Vibrational Absorption for Chemical
  Microscopy''}.
\newblock \emph{Nature} 1999; 399(6732): 134--137.
\newblock \doi{10.1038/20154}.
\newblock \urlprefix\url{https://doi.org/10.1038/20154}.

\bibitem{Taubner}
Taubner T, Hillenbrand R and Keilmann F.
\newblock {``Nanoscale Polymer Recognition by Spectral Signature in Scattering
  Infrared Near-Field Microscopy''}.
\newblock \emph{Appl~Phys~Lett} 2004; 85(21): 5064--5066.
\newblock \doi{10.1063/1.1827334}.
\newblock \urlprefix\url{https://doi.org/10.1063/1.1827334}.

\bibitem{Furstenberg_RSI06}
Furstenberg R, Soares JA and White JO.
\newblock {``Apparatus for the Imaging of Infrared Photoluminescence,
  Transmittance, and Phototransmittance with High Spatial and Spectral
  Resolutions''}.
\newblock \emph{Rev~Sci~Instrum} 2006; 77(7): 073101.
\newblock \doi{10.1063/1.2214931}.
\newblock \urlprefix\url{https://doi.org/10.1063/1.2214931}.

\bibitem{Ferrieu_RSI89}
Ferrieu F.
\newblock {``Infrared Spectroscopic Ellipsometry Using a {F}ourier Transform
  Infrared Spectrometer: {S}ome Applications in Thin‐Film
  Characterization''}.
\newblock \emph{Rev~Sci~Instrum} 1989; 60(10): 3212--3216.
\newblock \doi{10.1063/1.1140554}.
\newblock \urlprefix\url{https://doi.org/10.1063/1.1140554}.

\bibitem{Canillas_RSI93}
Canillas A, Pascual E and Dr{\'e}villon B.
\newblock {``Phase-Modulated Ellipsometer Using a {F}ourier Transform Infrared
  Spectrometer for Real Time Applications''}.
\newblock \emph{Rev~Sci~Instrum} 1993; 64(8): 2153--2159.
\newblock \doi{10.1063/1.1143953}.
\newblock \urlprefix\url{https://doi.org/10.1063/1.1143953}.

\bibitem{Kazarian_ApplSpctr07}
Chan KLA and Kazarian SG.
\newblock {``Attenuated Total Reflection Fourier Transform Infrared Imaging
  with Variable Angles of Incidence: A Three-Dimensional Profiling of
  Heterogeneous Materials''}.
\newblock \emph{Appl~Spectrosc} 2007; 61(1): 48--54.
\newblock \doi{10.1366/000370207779701415}.
\newblock \urlprefix\url{https://doi.org/10.1366/000370207779701415}.

\bibitem{Kazarian_ApplSpctr08}
Chan KLA, Tay FH, Poulter G et~al.
\newblock {``Chemical Imaging with Variable Angles of Incidence Using a Diamond
  Attenuated Total Reflection Accessory''}.
\newblock \emph{Appl~Spectrosc} 2008; 62(10): 1102--1107.
\newblock \doi{10.1366/000370208786049222}.
\newblock \urlprefix\url{https://doi.org/10.1366/000370208786049222}.

\bibitem{Kazarian_Langmuir10}
Frosch T, Chan KLA, Wong HC et~al.
\newblock {``Nondestructive Three-Dimensional Analysis of Layered Polymer
  Structures with Chemical Imaging''}.
\newblock \emph{Langmuir} 2010; 26(24): 19027--19032.
\newblock \doi{10.1021/la103683h}.
\newblock \urlprefix\url{https://doi.org/10.1021/la103683h}.

\bibitem{Kazarian_ApplSpctr15}
Wrobel TP, Vichi A, Baranska M et~al.
\newblock {``Micro-Attenuated Total Reflection Fourier Transform Infrared
  (Micro ATR FT-IR) Spectroscopic Imaging with Variable Angles of Incidence''}.
\newblock \emph{Appl~Spectrosc} 2015; 69(10): 1170--1174.
\newblock \doi{10.1366/15-07963}.
\newblock \urlprefix\url{https://doi.org/10.1366/15-07963}.

\bibitem{Kazarian_Analyst19}
Song CL and Kazarian SG.
\newblock {``Three-Dimensional Depth Profiling of Prostate Tissue by Micro
  ATR-FTIR Spectroscopic Imaging with Variable Angles of Incidence''}.
\newblock \emph{Analyst} 2019; 144: 2954--2964.
\newblock \doi{10.1039/C8AN01929K}.
\newblock \urlprefix\url{http://dx.doi.org/10.1039/C8AN01929K}.

\bibitem{Yao_OEX20}
Yao Y, Ikeda N, Kuroda T et~al.
\newblock {``Mid-IR {D}irac-Cone Dispersion Relation Materialized in {SOI}
  Photonic Crystal Slabs''}.
\newblock \emph{Opt~Express} 2020; 28(3): 4194--4203.
\newblock \doi{10.1364/OE.381996}.
\newblock \urlprefix\url{https://doi.org/10.1364/OE.381996}.

\bibitem{Sakoda_OEX12_1}
Sakoda K.
\newblock {``Dirac Cone in Two- and Three-Dimensional Metamaterials''}.
\newblock \emph{Opt~Express} 2012; 20(4): 3898--3917.
\newblock \doi{10.1364/OE.20.003898}.
\newblock \urlprefix\url{https://doi.org/10.1364/OE.20.003898}.

\bibitem{Sakoda_OEX12_2}
Sakoda K.
\newblock {``Proof of the Universality of Mode Symmetries in Creating Photonic
  {Dirac} Cones''}.
\newblock \emph{Opt~Express} 2012; 20(22): 25181--25194.
\newblock \doi{10.1364/OE.20.025181}.
\newblock \urlprefix\url{https://doi.org/10.1364/OE.20.025181}.

\bibitem{Williams_PTL12}
{Williams} DM, {Groom} KM, {Stevens} BJ et~al.
\newblock {``Epitaxially Regrown {GaAs}-Based Photonic Crystal Surface-Emitting
  Laser''}.
\newblock \emph{IEEE Photonics Technol~Lett} 2012; 24(11): 966--968.
\newblock \doi{10.1109/LPT.2012.2191400}.
\newblock \urlprefix\url{https://doi.org/10.1109/LPT.2012.2191400}.

\bibitem{Taylor_JSTQE12}
{Taylor} RJE, {Williams} DM, {Childs} DTD et~al.
\newblock ``{All-Semiconductor Photonic Crystal Surface-Emitting Lasers Based
  on Epitaxial Regrowth''}.
\newblock \emph{IEEE J Sel Top Quantum Electron} 2013; 19(4): 4900407.
\newblock \doi{10.1109/JSTQE.2013.2249293}.
\newblock \urlprefix\url{https://doi.org/10.1109/JSTQE.2013.2249293}.

\end{thebibliography}

%\input{anglRslvd_ApplSptrscpy_20200323.bbl}

%\begin{thebibliography}{99}
%\bibitem[Kopka and Daly(2003)]{R1}
%Kopka~H and Daly~PW (2003) \textit{A Guide to \LaTeX}, 4th~edn.
%Addison-Wesley.

%\bibitem[Lamport(1994)]{R2}
%Lamport~L (1994) \textit{\LaTeX: a Document Preparation System},
%2nd~edn. Addison-Wesley.

%\bibitem[Mittelbach and Goossens(2004)]{R3}
%Mittelbach~F and Goossens~M (2004) \textit{The \LaTeX\ Companion},
%2nd~edn. Addison-Wesley.

%\end{thebibliography}

\end{document}